SJANG L. TEN HAGEN*

# How "Facts" Shaped Modern Disciplines: The Fluid Concept of Fact and the Common Origins of German Physics and Historiography

**ABSTRACT**

This history of the concept of fact reveals that the fact-oriented practices of German physicists and historians derived from common origins. The concept of fact became part of the German language remarkably late. It gained momentum only toward the end of the eighteenth century. I show that the concept of fact emerged as part of a historical knowledge tradition, which comprised both human and natural empirical study. Around 1800, parts of this tradition, including the concept of fact, were integrated into the epistemological basis of several emerging disciplines, including physics and historiography. During this process of discipline formation, the concept of fact remained fluid. I reveal this fluidity by unearthing different interpretations and roles of facts in different German contexts around 1800. I demonstrate how a fact-based epistemology emerged at the University of Göttingen in the late eighteenth century, by focusing on universal historian August Ludwig Schlözer and the experimentalist Georg Christoph Lichtenberg. In a time of scientific and political revolutions, they regarded facts as eternal knowledge, contrasting them with short-lived theories and speculations. Remarkably, Schlözer and Lichtenberg construed facts as the basis of *Wissenschaft*, but not as *Wissenschaft* itself. Only after 1800, empirically minded German physicists and historians granted facts self-contained value. As physics and historiography became institutionalized at German universities, the concept of fact acquired different interpretations in different disciplinary settings. These related to fact-oriented research practices, such as precision measurement in physics and source criticism in historiography.

*Institute for Theoretical Physics Amsterdam, Vossius Center for the History of Humanities and Sciences, University of Amsterdam; Science Park 904, 1098 XH, Amsterdam, The Netherlands; s.l.tenhagen@uva.nl

## INTRODUCTION

Today, due to the ubiquity of "fake news" and "alternative facts," it seems to be increasingly difficult to distinguish between fact and fiction.[1] For sure, this is an alarming development. But for some historians, philosophers, and sociologists of knowledge, the current circumstances offer at least one advantage: the rise of the so-called post-truth society has put facts under a magnifying glass, and made their work, which analyzes the processes involved in the making of facts, even more urgent. Yet, one fundamental issue has remained underexposed in the literature on the dynamics of factual knowledge:[2] what actually is a "fact"?

Although more urgent than ever, it turns out that this question is difficult to answer. From a historical point of view, this is because there have existed many different interpretations of the concept of fact over the past few centuries. It would be misleading to assume that the definition of a fact is etched in stone. Indeed, the fact is a relatively recent notion, and it has experienced a tumultuous history.[3] Like "probability" and "objectivity,"[4] it has never been self-evident, even though today these epistemological categories may produce, to quote Ian Hacking, a "feeling of inevitability."[5]

---

1. The phrase "alternative facts" was introduced by Kellyane Conway in 2017 and then cultivated by Donald Trump and his entourage. Jon Swaine, "Donald Trump's team defends 'alternative facts' after widespread protests," *The Guardian*, 23 Jan 2017; https://www.theguardian.com/us-news/2017/jan/22/donald-trump-kellyanne-conway-inauguration-alternative-facts (accessed 18 Oct 2018).

2. Two classic examples are: Bruno Latour and Steve Woolgar, *Laboratory Life: The Social Construction of Scientific Facts* (London: Sage Publications, 1979); Steven Shapin and Simon Schaffer, *Leviathan and the Air-pump* (Princeton, NJ: Princeton University Press, 1985). A more recent example is: Peter Howlett and Mary S. Morgan, eds., *How Well Do Facts Travel?* (New York: Cambridge University Press, 2011).

3. On the early modern origins of the concept of fact: Lorraine Daston, "Marvelous Facts and Miraculous Evidence in Early Modern Europe," *Critical Inquiry* 18, no. 1 (1991): 93–124; Mary Poovey, *A History of the Modern Fact: Problems of Knowledge in the Sciences of Wealth and Society* (Chicago: University of Chicago Press, 1998); Barbara J. Shapiro, *A Culture of Fact: England, 1550–1720* (New York: Cornell University Press, 2000). On the use of the concept of fact in later periods: e.g., Theodore M. Porter, "Statistical and Social Facts from Quetelet to Durkheim," *Sociological Perspectives* 38, no. 1 (1995): 15–26; Jan Frercks, "Demonstrating the Facticity of Facts: University Lectures and Chemistry as a Science in Germany around 1800," *Ambix* 57, no. 1 (2010): 64–83; Maria Pia Donato, "A Science of Facts? Classifying and Using Records in the French Imperial Archives under Napoleon," *History of Humanities* 2, no. 1 (2017): 79–100.

4. Ian Hacking, *The Emergence of Probability* (Cambridge: Cambridge University Press, 1975); Lorraine Daston and Peter Galison, *Objectivity* (New York: Zone Books, 2007).

5. Ian Hacking, *Historical Ontology* (Cambridge, MA: Harvard University Press, 2002), 21.

This misleading feeling of inevitability might have also influenced those uncovering the social processes by which facts are made, such as Bruno Latour. In an essay published in 2004, Latour admitted that he had previously been "accepting much too uncritically what matters of fact were."[6] One finds similar observations in the work of historical epistemologists such as Lorraine Daston.[7] "Although historians and philosophers of science have worried that facts may be 'contaminated' by theory or 'constructed' by society," Daston has noted, "the *category* of the factual has remained curiously unanalyzed...The concept of what kind of thing or event qualifies as a scientific fact, and when and why it does, has escaped investigation."[8]

In this article, I study the origins and transformations of the concept of fact in German academic culture around 1800. This era was crucial for the emergence of modern *Wissenschaft*, which comprised a system of disciplines we now consider to belong to either the sciences or the humanities.[9] As it turns out, facts played a significant role in this development. I highlight this role by focusing on the increasing importance of facts in two disciplines that took shape around 1800: physics and historiography.

Historians of science have referred to the period around 1800 as the "great transition."[10] During this transitional phase, the hierarchically related

---

6. Bruno Latour, "Why Has Critique Run out of Steam? From Matters of Fact to Matters of Concern," *Critical inquiry* 30, no. 2 (2004): 225–48, on 231.

7. In Hacking's words, the subgenre of historical epistemology—sometimes labeled "historical meta-epistemology" or "historical ontology"—studies "epistemological concepts...as objects that evolve and mutate." Hacking, *Historical Ontology* (ref. 5), 8–9. On historical epistemology, see also: Lorraine Daston, "Historical Epistemology," in *Questions of Evidence: Proof, Practice, and Persuasion across the Disciplines*, ed. James Chandler, Arnold I. Davidson, and Harry D. Harootunian (Chicago: University of Chicago Press, 1994), 282–89; Hans-Jörg Rheinberger, *On Historicizing Epistemology: An Essay* (Stanford, CA: Stanford University Press, 2010); Omar W. Nasim, "Was ist historische Epistemologie?," in *Nach Feierabend*, ed. David Gugerli et al., (Zürich: Diaphanes, 2013), 123–44.

8. Lorraine Daston, "Description by Omission: Nature Enlightened and Obscured," in *Regimes of Description: In the Archive of the Eighteenth Century*, ed. John Bender and Michael Marrinan (Stanford, CA: Stanford University Press, 2005), 11–24, on 12–13.

9. The German term *Wissenschaft* can refer either to the whole of natural and human knowledge enterprises, or to the body of knowledge they produce. During the nineteenth century, German physicists and historians both labelled their subjects as *wissenschaftlich*. It would thus be misleading to consider *Wissenschaft* as a direct translation of the English term "science," which is usually taken to exclude disciplines such as history and philology and the knowledge these disciplines produce.

10. Originally proposed in the 1980s, there have been recent attempts to start reusing the term: R. Steven Turner, "The Great Transition and the Social Patterns of German Science," *Minerva* knowledge traditions of mathematics, natural philosophy, and natural history gave way to a non-hierarchical system of disciplines.[11] The period around 1800 was of crucial importance for the development of the modern humanities as well. The study of language, culture, and history experienced major shifts at the turn of the nineteenth century, on both a conceptual and an institutional level.[12] For example, it was the formative period of German historicism.[13] Furthermore, disciplines like history and linguistics became institutionalized at German universities.[14]

Although the formation of modern *Wissenschaft* in the decades around 1800 has been studied fairly extensively, there is still a lot to gain.[15] For one thing, the literature about the period disintegrates into two largely separated historiographies: one focuses on the sciences, the other on the humanities.[16] Whereas such a dualistic evaluation may seem logical from a contemporary point of view, it fails to do justice to the reality of scholarship at the time.

Before 1800, few German intellectuals focused on either human or natural study exclusively. For a considerable part of the nineteenth century, moreover, the categories of science (*Naturwissenschaften*) and humanities (*Geisteswissenschaften*) did not exist.[17] Because of the entanglement of human and natural study in the German-speaking lands in the decades around 1800, it is important to consider these realms of knowledge in relation to one another.[18]

My focus on the history of the concept of fact demands such integration.[19] Like recently written histories of "data," this history of the fact draws attention to "relationships, convergences, and contingent historical developments that can be obscured following more traditional areas of focus on individual disciplines," thus highlighting "interactions between ostensibly distinct knowledge traditions."[20] In particular, this history of the fact reveals an essential link between the disciplines of physics and history. In what follows, I show that the fact-oriented epistemologies and practices of German physicists and historians, which gained momentum over the course of the nineteenth century, derived from common origins. The concept of fact in Germany initially existed as part of a so-called historical knowledge tradition, which comprised *any* kind of empirical study. Around 1800, elements of this historical tradition, including the concept of fact, became part of the epistemological foundations of the emerging disciplines of history and physics. I argue that, as part of this transformation, the roles and interpretations assigned to facts transformed as well.

---

17. Even after these categories had emerged, there were many interconnections between humanities and science. For example, plant physiologists and paleographers collaborated to date old manuscripts, physicists and astronomers analyzing spectra were inspired by deciphering practices in Egyptology, and historicism was not confined to the historical sciences, but extended into the life sciences. Josephine Musil-Gutsch, "On the Same Page: Investigating Material Remains of Science and Humanities' Knowledge-Producing Practices," *History of Humanities* (forthcoming); Matthias Dörries, "Heinrich Kayser as Philologist of Physics," *Historical Studies in the Physical and Biological Sciences* 26, no. 1 (1995): 1–33; Marianne Klemun, "Historismus/Historismen—Geschichtliches und Naturkundliches: Identität—Episteme—Praktiken," in *Wissenschaftliche Forschung in Österreich 1800–1900: Spezialisierung, Organisation, Praxis*, ed. Christine Ottner, Gerhard Holzer, and Petra Svatek (Göttingen: V&R unipress, 2015), 17–42.

18. Recently, there have been several calls to study the history of humanities and science in an integrated manner: Rens Bod and Julia Kursell, "Focus: The History of Humanities and the History of Science," *Isis* 106, no. 2 (2015): 337–90; Fabian Krämer, "Forum: The Two Cultures Revisited; The Sciences and the Humanities in a *Longue Durée* Perspective," *History of Humanities* 3, no. 1 (2018): 5–88; Rens Bod, Jeroen van Dongen, Sjang L. ten Hagen, Bart Karstens, and Emma Mojet, "The Flow of Cognitive Goods: A Novel Framework for the Historical Study of Epistemic Transfer," *Isis* (forthcoming).

19. Mary Poovey has argued that her focus on the history of the concept of fact enabled her "to expose the connections between knowledge projects as different as rhetoric, natural philosophy, moral philosophy, and early versions of the modern social sciences." Poovey, *History of Modern Fact* (ref. 3), xiv–xv. On the general relation between conceptual history and interdisciplinarity: Irmline Veit-Brause, "The Interdisciplinarity of History of Concepts—A bridge between disciplines," *History of Concepts Newsletter* 6 (2003): 8–13; Ernst Müller, "Introduction: Interdisciplinary Concepts and their Political Significance," *Contributions to the History of Concepts* 6, no. 2 (2011): 42–52.

20. David Sepkoski, "Data in Time: Statistics, Natural History, and the Visualization of Temporal Data," *Historical Studies in the Natural Sciences* 48, no. 5 (2018), 581–93, on 581. On the history of "data," see the contributions in: Soraya de Chadarevian and Theodore M. Porter, "Special Issue: Histories of Data and the Database," *Historical Studies in the Natural Sciences* 48, no. 5 (2018), 549–684; Elena Aronova, Christine von Oertzen, and David Sepkoski, eds., *Data Histories*, *Osiris* 32 (2017).

To sum up, my aim in this paper is twofold. First, I aim to enrich the history of the concept of fact by examining its emergence and transformations in the German context around 1800. Second, I aim to uncover connections between the formation histories of physics and historiography by drawing parallels between the emergence of fact-oriented methods in these disciplines.

In the next section, I begin by explaining how the concept of fact entered German scholarly language.[21] The core of this study follows in the subsequent sections, which focus on the late-eighteenth-century University of Göttingen. First, I explain why this is a relevant context to study the dynamics of the great transition. Subsequently, I examine the rise of a fact-based epistemology among Göttingen scholars with seemingly diverse yet epistemologically related backgrounds, including *Univeral-Historie* and *Naturlehre*. In the final parts of this article, I examine how the still fluid concept of fact became part of the disciplinary vocabularies of physicists and historians in the early nineteenth century. My focus then shifts from Göttingen toward main nodes of discipline formation in the German-speaking lands after 1800. I argue that the concept of fact experienced a profound transformation in these contexts, which was an important factor in the materialization of physics and historiography as autonomous academic disciplines.

## THE 18TH-CENTURY ORIGINS OF THE *WÖRTLEIN THATSACHE*

The origins of the early modern usage of the concept of fact lie in sixteenth-century England. The concept was first used in English law and human history. For sixteenth-century English lawyers and historians, facts, or "matters of fact," were alleged acts or deeds of which the occurrence was in contention. So, a fact could be untrue. When scholars started to apply the concept of fact to study nature, its meaning changed. For seventeenth-century natural philosophers, facts became empirically established truths. The emergence of the English "culture of fact" has been studied thoroughly.[22] But much less is known about the introduction of the fact to other languages.

---

21. Unfortunately, the history of the fact in German public spheres falls beyond the scope of this study. Yet it is most likely that the term "fact," like the term *Naturwissenschaft*, was a co-product of learned and public environments. On the latter example: Denise Phillips, *Acolytes of Nature: Defining Natural Science in Germany, 1770–1850* (Chicago: The University of Chicago Press, 2011).

Histories of the fact in different European cultures were intertwined, since these cultures all developed from the Latin early modern knowledge tradition. The Latin term *factum* was a common point of reference for scholars working in France, England, and Germany.[23] Remarkably, the German language incorporated the concept rather late when compared to the English, but also to the French. An explanation for this may be that Latin remained the dominant language among intellectuals in the German states until the beginning of the eighteenth century.[24] German only gained prestige from the beginning of the eighteenth century onward, when prominent German philosophers such as Christian Wolff (1679–1754) began to publish in the vernacular.[25] With this development came a need for linguistic innovations to accommodate for concepts that had until then only existed in other languages. The fact was one of such examples. The term *factum* already existed in Latin texts by German scholars, and was usually interpreted as a deed or event.[26] From the mid-eighteenth century onward, *Factum* (or *Faktum*) began to be used as a German word as well, and was interpreted the same as its Latin precedent.[27]

To some extent, the introduction of the concept of fact in German culture can thus be understood as part of a conversion of the conceptual framework of knowledge from Latin to German: *factum* became *Factum*. Yet, this is only part of the story. A crucial moment in the German history of the fact was in 1756, when the term *Thatsache* was coined by the Scottish-German pastor Johann Joachim Spalding (1714–1804).[28] The term was a translation of a term used by the English theologian Joseph Butler's (1692–1752): "matter of fact."[29] In the 1770s and 1780s, *Thatsachen* played a major role in debates about Deism in German theology, where they were defined as historically attested biblical acts, actions, and events. In this particular context, facts were not regarded as true, but only as probable.[30]

---

22. Shapiro, *Culture of Fact* (ref. 3).
23. Lorraine Daston, "Baconsche Tatsachen," *Rechtsgeschichte* 1, no. 1 (2002): 36–55, on 39.
24. Michael D. Gordin, *Scientific Babel* (Chicago: The University of Chicago Press, 2015), 42; Phillips, *Acolytes of Nature* (ref. 21), 53–57.
25. Wolgang Walter Menzel, *Vernakuläre Wissenschaft, Christian Wolffs Bedeutung für die Herausbildung und Durchsetzung des Deutschen als Wissenschaftssprache* (Tübingen: Niemeyer, 1996).
26. A renowned early eighteenth-century universal lexicon defined the term *factum* as a deed or event. In German: "eine That, das geschehene Ding, oder eine Geschichte, das Werck, die Verrichtung, der Verlauff eines ergangenen Handels." Johann H. Zedler, *Grosses Vollständiges Universallexicon Aller Wissenschaften und Künste*, vol. 9, (Leipzig/Halle, 1735), 65–66.
27. Hans Schulz and Otto Basler, *Deutsches Fremdwörterbuch*, vol. 5 (Berlin: De Gruyter, 2004), 658–60.

In the late eighteenth century, the co-existence of the terms *Factum* and *Thatsache* in the German language made for a complex terminological picture. What is more, the precise meaning and function of these terms remained an open issue. An astute observation by the prominent German writer and theologian Gotthold Ephraim Lessing (1729–1781) illustrates the point. In 1781, Lessing noted how suddenly and dramatically the use of the *Wörtlein Thatsache* had increased: "I rightly say: *little word*, while it is still so young. I very well remember the time when nobody used it. I do not know how this little word [*Thatsache*] has become so popular that in some literature one cannot turn a page without stumbling upon a [*Thatsache*]." Lessing also discussed the term *Factum*, which he identified as related but of which he thought the meaning was already clear: "[*Facta*] and [*Faits*] are just [*geschehene Dinge*, *Begebenheiten*, *Thaten*, *Ereignisse*, *Vorfälle*] of which the historical certainty is as great as historical certainty can be."[31] The exact meanings of and relations between these increasingly popular terms, especially *Thatsache*, were unclear to Lessing. His remarks reveal that the concept of fact, in its multiple linguistic guises, was regarded as both novel and fluid in Germany in the late eighteenth century.

The use of the terms *Factum*, *Faktum*, and *Thatsache* increased exponentially in the final decades of the eighteenth century (Fig. 1). Simultaneously, the

---

28. Karlfried Gründer, *Historisches Wörterbuch der Philosophie*, vol. 10 (Basel: Schwabe Verlag, 1998), 910–16. Around 1900, the common spelling of the word became *Tatsache*.

29. The theological-historical origins of the concept of fact in Germany thus interlinked with the English context. Shapiro, *Culture of Fact* (ref. 3), 168–88.

30. Reinhart Staats, "Der Theologiegeschichtliche Hintergrund des Begriffes 'Tatsache'," *Zeitschrift für Theologie und Kirche* 70, no. 3 (1973): 316–45.

31. Gotthold E. Lessing, "Über das Wörtlein Thatsache," in *Sämmtliche Schriften*, vol. 11, ed. Karl Lachmann (Berlin: Voß, 1839), 645. Translation of German quotes into English are mine, unless stated otherwise.

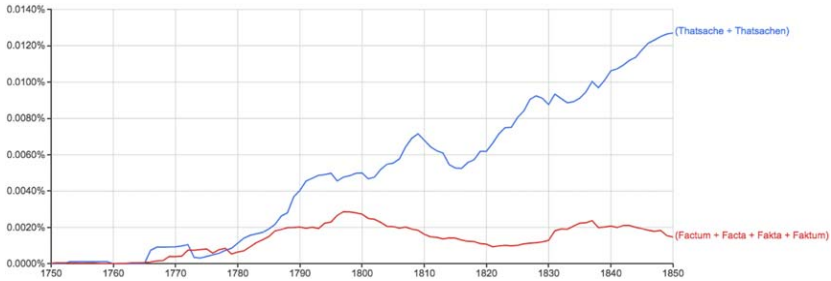

**FIG. 1.** *Factum/Facta* and *Thatsache(n)* in German texts. Relative summed frequencies of "Thatsache/Thatsachen" and "Factum/Faktum/Facta/Factum" in German texts, as obtained on 13 February 2018 via Google Books Ngram Viewer, using a smoothing of 3. Given the evident limits of the Google Books text corpus for historical research, particularly in the period studied, this plot should be interpreted not too literally. Nevertheless, it does give a clear indication both of the late eighteenth-century origin of the "fact" in Germany and the co-existence of *Factum* and *Thatsache* in this period. On Google Books as a tool for research, see: Eitan Adam Pechenick, Christopher M. Danforth, and Peter Sheridan Dodds, "Characterizing the Google Books corpus: Strong limits to inferences of socio-cultural and linguistic evolution," *PloS one* 10, no. 10 (2015): e0137041.

Latin words *datum* and *data* came into wider circulation.[32] These and other German terms, such as *Begebenheit*, all related to the concept of fact that gradually emerged in Germany around 1800.[33] Ultimately, however, *Thatsache* became a much more popular term than its Latinist counterparts.

How can the growing popularity of these terms be explained? And which transformations occurred as the concept of fact spread through German learned culture in different verbal forms? In the following, I aim to answer these questions by focusing on the University of Göttingen. I discuss how and why "facts" became important in Göttingen physics and historiography in the late eighteenth century. I also show that, while this happened, these still interrelated fields of study gradually began to differentiate.

---

32. In 1768, Herder used *factum* and *datum* interchangeably with *Begebenheit*. Johann G. Herder, "Älteres kritisches Wäldchen," in *Schriften zur Ästhetik und Literatur 1767–1781*, ed. Gunter E. Grimm (Frankfurt am Main: Deutsche Klassiker Verlag, 1993), 11–56, on 11–12.

33. On the emergence of the concept of fact in late eighteenth-century German philosophy and literature: Paul Ziche, "'Höherer' Empirismus: Passive Wissenschaft, Letzte Tatsachen und Experimentelle Philosophie bei FWJ Schelling," in *Philosophie und Wissenschaft / Philosophy and Science*, ed. Jürgen Stolzenberg and Fred Rush (Berlin: De Gruyter, 2011), 165–84; Juliane Vogel, "Die Kürze des Faktums: Textökonomien des Wirklichen um 1800," *Deutsche Vierteljahrsschrift für Literaturwissenschaft und Geistesgeschichte* 89, no. 3 (2015): 297–306; Johannes Lehmann, "Faktum, Anekdote, Gerücht: Zur Begriffsgeschichte der 'Thatsache' und 'Kleists Berliner Abendblättern'," *Deutsche Vierteljahrsschrift für Literaturwissenschaft und Geistesgeschichte* 89, no. 3 (2015): 307–22; Jocelyn Holland, "Facts are What One Makes of Them: Constructing the *Faktum* in the Enlightenment and Early German Romanticism," in *Fact and Fiction: Literary and Scientific Cultures in Germany and Britain*, ed. Christine Lehleiter (Toronto: University of Toronto Press, 2016), 33–49.

## THE UNIVERSITY OF GÖTTINGEN BEFORE 1800

By the end of the eighteenth century, the city of Göttingen in the Kingdom of Hannover had developed into the main knowledge hub of the Holy Roman Empire. The city's young and vibrant *Georg-August-Universität*, founded in 1737, was populated by an internationally oriented and renowned intellectual community. The scholarly environment in Göttingen also featured a world-famous library, leading journals, and a prominent Academy.[34]

In the Göttingen context, the so-called fourth faculty of philosophy experienced a boost, and gained independence with respect to the previously dominant faculties of law, medicine, and theology. The philosophical faculty accommodated relatively independent institutional environments for a range of subject areas, for example, by implementing increasingly specialized professorial chairs. The institutional structure of Göttingen's university functioned as an example for nineteenth-century reforms at other German universities, among them, the university in Berlin.[35] For these reasons, the late-eighteenth-century Göttingen context is a suitable one for studying the transformations that led to the establishment of the modern system of knowledge disciplines.

As a result of the foundation of specialized professorial chairs and disciplinary institutes, physics and historiography started to develop within separate institutional contexts in late-eighteenth-century Göttingen. With regard to historiography, an important development was the foundation of the specialized Institute for Historical Sciences in 1764. Martin Gierl has argued that this institute linked "systematic research, training of historians, recording of sources and institutionalization of the discipline," and hence was "an institutional signpost in the development of modern historiography."[36] In the following, I discuss the role and importance of facts in the so-called Göttingen School of History, paying specific attention to its authoritative members, Johann Christoph Gatterer (1727–1799) and August Ludwig Schlözer (1735–1809). Subsequently, I focus on Georg Christoph Lichtenberg (1742–1799), who played a significant role in the development of German experimental physics.

---

34. Dirk van Miert, "Structuring the History of Knowledge in an Age of Transition: The Göttingen Geschichte between Historia Literaria and the Rise of the Disciplines," *History of Humanities* 2, no. 2 (2017): 389–416, on 392.

35. Rudolf Stichweh, "The Sociology of Scientific Disciplines: On the Genesis and Stability of the Disciplinary Structure of Modern Science," *Science in Context* 5, no. 1 (1992): 3–15, on 9.

Although dynamics of differentiation had gotten into gear, neither physics nor history functioned as autonomous disciplines in Göttingen or elsewhere in the German-speaking states around 1800. Physics (*Physik* or *Experimentalphysik*), like chemistry (*Chemie*), existed as a combination of elements from natural philosophy (*Naturlehre*), natural history (*Naturgeschichte*), and applied mathematics (*Angewandte Mathematik*).[37] When experimental physics and chemistry came to be seen as independent areas of research, new classifications of knowledge started to interfere with older ones; different definitions of *Naturlehre*, *Physik*, and *Chemie* existed alongside each other. The uncrystallized status of "physics" in Germany and the shifting significance of *Naturlehre* and *Physik* around 1800 implies that it is not without problems to refer to Lichtenberg and his contemporaries as "physicists." Indeed, Lichtenberg's intellectual pursuits extended over a range of subject areas, which remained fundamentally intertwined.

The same applies to the Göttingen "historians" I have just mentioned. Although Gatterer and Schlözer aimed to establish autonomous ground for history, their scholarly enterprise was strongly affiliated with philosophical, philological, and statistical practice.[38] Moreover, both Gatterer and Schlözer remained oriented to the content and methods of natural research.

So, even though scholars like Lichtenberg and Schlözer sometimes used seemingly modern labels such as "physics" and "history," these labels had different meanings than today. Furthermore, even though these scholars

36. Martin Gierl, "Change of Paradigm as a Squabble between Institutions," in *Scholars in Action: The Practice of Knowledge and the Figure of the Savant in the 18th Century*, ed. André Holenstein, Hubert Steinke, and Martin Stuber (Leiden: Brill, 2013), 286–87.

37. Stichweh, *Zur Entstehung wissenschaftlicher Disziplinen* (ref. 11), 14–38.

38. Schlözer's use of tables has been identified as "data-driven" practice in the context of his work as a statistician, which illustrates his multilateral profile as a researcher. David Sepkoski and Marco Tamborini, "'An Image of Science': Cameralism, Statistics, and the Visual Language of Natural History in the Nineteenth Century," *Historical Studies in the Natural Sciences* 48, no. 1 (2018): 56–109, on 68; Christine von Oertzen, "Datafication and Spatial Visualization in Nineteenth-Century Census Statistics," *Historical Studies in the Natural Sciences* 48, no. 5 (2018): 568–80, on 570.

played significant roles in the differentiation of physics and history, they were not part of disciplinary frameworks. Their use and interpretation of the category of fact, which corresponded on crucial points, reveals the strong connection between the scholarly ideals and practices of the "physicist" Lichtenberg and the "historian" Schlözer.

## FACTS IN THE GÖTTINGEN SCHOOL OF HISTORY

August Ludwig Schlözer and Johann Christoph Gattterer have been portrayed as founding figures of German professional historiography.[39] Both aimed to free historical study from the burden of law and theology by claiming an independent place for the field in the philosophical faculty of Göttingen's university. Furthermore, they promoted their enterprise in the public sphere.[40]

The methodologies of Gatterer and Schlözer converged on many points. They differed, however, in what they considered the basic units of universal history. For Gatterer, these were "events" (*Begebenheiten*). In 1792, for example, he argued that "events are the actual subject matter of history: from the immense quantity of them, one selects the most remarkable, proves their reality from the sources, and narrates them in relation to one another."[41] Schlözer agreed with Gatterer that real events could be building blocks of sound historical research. However, he specified the units of history beyond a mere event by introducing the concept of fact. This innovation enabled Schlözer to label non-events like "the kings of Mycenae," together with events such as the arrival of pepper in Europe and the Battle of Carchemisch, as the basic units of history. These were all facts.[42] On many occasions, Schlözer used "events" and "facts" interchangeably.

---

39. Peter H. Reill, *The German Enlightenment and the Rise of Historicism* (Berkeley: University of California Press, 1975); Martin Peters, *Altes Reich und Europa: der Historiker, Statistiker und Publizist August Ludwig (v.) Schlözer (1735–1809)* (Münster: LIT Verlag, 2003); Martin Gierl, *Geschichte als präzisierte Wissenschaft: Johann Christoph Gatterer und die Historiographie des 18. Jahrhunderts im ganzen Umfang*, Fundamenta Historica, vol. 4 (Stuttgart: Frommann-Holzboog, 2012).

40. Schlözer's lectures at the University of Göttingen attracted hundreds of students, and his periodic journals were read by thousands. Justin Stagl, *A History of Curiosity* (Chur: Harwood, 1995), 246.

41. As quoted by Claudia Schrapel, *Johann Dominicus Fiorillo* (Zürich: Hildesheim, 2004), 175.

42. August L. Schlözer, *Vorstellung seiner Universal-Historie*, vol. 2 (Göttingen/Gotha, 1773), 253.

In the first edition of Schlözer's *Vorstellung seiner Universal-Historie* (1772), he argued that the aim of universal history, or world history (*Weltgeschichte*), was to present as many facts as possible and to present them in relation to one another: "the best world history is the one that contains the most expediently selected [*Facta*]."[43] The more facts a historical work presented, the better, Schlözer claimed. But he also believed that one should not stick to the mere collection of facts: "The best method of world history is the one that makes it easiest for its student to understand the coherence of all these [*Factorum*]."[44] Schlözer cherished the Enlightenment ideals of universality: he aimed to present a unified history of humanity, spanning all centuries and continents, and connecting the past to the present.[45] In order to reach such a universal picture, Schlözer declared it to be the ultimate aim of history to combine and to organize facts "synthetically."[46] For this, Schlözer used tables, which ordered relevant historical facts either according to time period or according to people (Fig. 2).[47] Following Schlözer, the collection of self-explanatory facts or events formed the empirical and certain basis of historical science, whereas only the ordering of them could lead to divergent pictures. To illustrate this, Schlözer drew a parallel with mosaic artists, who can build entirely different portraits from the very same collection of stones.[48]

Schlözer used different words to denote the concept of fact throughout his oeuvre. In the 1773 reissue of Schlözer's *Vorstellung seiner Universal-Historie*, he argued that the terms *Begebenheit* and *factum* were unfitting as the basic units of history. Hence, he proposed a new term: *Thatsatz*. He justified this linguistic innovation as follows: "May I use [*Thatsatz*] rather than the alien and inflexible [*factum historicum*]? [*Begebenheit*] does not always fit. In general, one must not avoid a little bit of neology if world history is to be presented [*scientifisch*]."[49] Alas, Schlözer did not further define his neologism; the precise meaning of *Thatsatz* follows only implicitly from his writings. Possibly, Schlözer proposed

---

43. August L. Schlözer, *Vorstellung seiner Universal-Historie*, vol. 1 (Göttingen/Gotha, 1772), 45.
44. Ibid., 45.
45. Ibid., 4.
46. Ibid., 96–98.
47. Schlözer's work contained "synchronistic" tables (ibid., 89–93) and "ethnographic" tables (ibid. 109–10). He argued that such tables form the basis of universal history: Schlözer, *Vorstellung*, vol. 2 (ref. 42), 248. On the use of tables in early modern historiography and their relation to facts: Benjamin Steiner, *Die Ordnung der Geschichte: Historische Tabellenwerke in der Frühen Neuzeit* (Köln: Böhlau Verlag, 2008), ff. 236.
48. Schlözer, *Vorstellung*, vol. 2 (ref. 42), 254.
49. Ibid., 235 (footnote).

**FIG. 2. Schlözer's fact-ordering tables.**
A fragment of one of Schlözer's *synchronistische Tabellen*, ordering historical facts synthetically, according to time period. *Source*: Schlözer, *Vorstellung 1* (ref. 43), 89, from Niedersächsische Staats- und Universitätsbibliothek Göttingen, 8 H UN II, 510:1 RARA.

the term (which may be translated as "factual proposition") to justify that the basic units of his universal history were often generalizations rather than individual events. Nonetheless, Schlözer's introduction of the novel term *Thatsatz*, which was never broadly adopted, underlines the ambiguities surrounding the concept of fact in the German context around 1800. In later writings, Schlözer alternately used the terms *Factum*, *Tatsatz*, *Tatsache* (both now without an "h"), as well as their plurals, in his description of proper historical method.[50] It seems

50. In 1785, Schlözer distinguished between three different roles in historical practice, using *Faktum*, *Tatsatz*, and *Tatsache* interchangeably: the *GeschichtSammler* collects as many facts as that Schlözer eventually considered these terms as synonyms, but it is difficult to determine at what point Schlözer adopted which one, and whether he interpreted them differently or not.

**Against Speculative History**

Schlözer's fact-based epistemology matched a broader empirical turn among Göttingen scholars. This turn was rooted in philological practice, particularly in the philological method of source criticism, which was practiced in the Göttingen philological seminar.[51] Because many history students attended, the seminar played a prominent role in making philological method a key part of historical practice. Until 1763, this successful teaching institute was being led by the theologically inclined scholar Johann David Michaelis (1717–1791). For Michaelis, who was Schlözer's mentor,[52] proper empirical history required knowledge of the facts. When Michaelis referred to facts, he emphasized, like Schlözer, that they were found in manuscripts and that they stood in sharp contrast with conjectures.[53]

Despite internal struggles, the Göttingen School's representatives, including historically minded philologists like Michaelis and philologically minded historians like Schlözer, collectively opposed speculative historiography, which they perceived as a threat to their empiricism. From the 1770s onward, they did so by insisting on the importance of facts and the associated method of source criticism. Their arrows were mostly aimed at French historians. For example, Schlözer at one point proclaimed that "critical method digs up the [*Facta*] from annals and monuments one by one (the Voltaires make them themselves, or at least color them)."[54]

---

possible from historical manuscripts, the *GeschichtForscher* critically investigates their authenticity, and the *GeschichtSchreiber* oversees all available and critically examined facts, and draws from them only the most relevant ones. August L. Schlözer, "Schlözer über die Geschichtsverfassung," in *Theoretiker der deutschen Aufklärungshistorie*, Fudamenta Historica 1.2, vol. 2, ed. Horst W. Blanke and Dirk Fleischer (Stuttgart: Fromman-Holzboog, 1990), 590–99.

51. William Clark, *Academic Charisma and the Origins of the Research University* (Chicago: The University of Chicago Press, 2006), 141–82.

52. Reill, *Rise of Historicism* (ref. 39), 84.

53. Johann D. Michaelis, *Einleitung in die göttlichen Schriften des Neuen Bundes*, 4th ed., vol. 1, (Göttingen: Vandenhoek, 1788), 574 and 794.

54. Schlözer, *Vorstellung*, vol. 1 (ref. 43), 45. On Gatterer and Schlözer's distaste for French historiography, see Reill, *Rise of Historicism* (ref. 39), 36–37.

Schlözer's polemic rhetoric was not only aimed at the French, but also at a prominent, German-speaking contemporary: Johann Gottfried Herder (1744–1803). In the beginning of the 1770s, Schlözer and Herder engaged in a fierce conflict about the proper historical method. The reason for the intellectual dispute had been a scathing review of Schlözer's 1772 *Universal-Historie* written by Herder in that same year.[55] The 1773 reissue of Schlözer's *Universal-Historie* was written as a direct response to Herder's review. Here, Schlözer blamed Herder for a lack of interest in "facts," distinguishing his own work from Herder's *Schmuck*. According to Schlözer, Herder, much unlike a proper historian, cared not about the facts themselves, but only about displaying them as gracefully as possible.[56]

Herder, in turn, disagreed not so much with Schlözer's emphasis on the importance of facts, but especially with his dedication to *Universal-Historie*, which was built on the assumptions of historical progress and universality. Herder rejected these assumptions, and insisted on contextualization and the principle of individuality instead.[57] All the same, Herder eventually buried the hatchet. In 1798, he even wrote an acclaiming review of one of Schlözer's later works, praising it for being "a book that develops and presents its merits from [*Thatsachen*]."[58] Herder's compliment illustrates that the standpoint that facts were the units of history had become increasingly common toward the end of the eighteenth century.

### A Blend of Human and Natural Study

I argued before that, despite early dynamics of disciplinary differentiation, research into natural and human history remained closely intertwined in Göttingen, at least until the late eighteenth century.[59] Schlözer and Gatterer commonly expressed the view that natural and human history formed a whole. For example, they both took the Linnaean classification system of natural history as a reference point while developing methods to order facts or events systematically.[60]

---

55. The review was published on 28 July: Johann G. Herder, review of *Vorstellung seiner Universal Historie*, by August L. Schlözer, *Frankfurter gelehrte Anzeigen* 60 (1772): 473–78. For more context, see: Justin Stagl, "Rationalism and Irrationalism in Early German Ethnology: The Controversy between Schlözer and Herder, 1772/73," *Anthropos*, 93, no. 4 (1998): 521–36.

56. Schlözer, *Vorstellung*, vol. 2 (ref. 42), 384.

57. Beiser, *Historicist Tradition* (ref. 13), 98–166.

58. As quoted by Peters, *Schlözer* (ref. 39), 404.

59. Historian of science John Heilbron has typified the prevalent spirit at the university as "a blend of history and natural science." John L. Heilbron, "Physics and Its History at Göttingen around 1800," in *Göttingen and the Development of the Natural Sciences*, ed. Nicolaas A. Rupke (Göttingen: Wallstein Verlag, 2002), 50–71, on 67.

Sometimes, Göttingen historians and naturalists were involved in the very same intellectual projects. In the 1790s, Johann Gottfried Eichhorn (1752–1827) initiated a laborious enterprise called the *Geschichte der Künste und Wissenschaften*, which appeared in 61 volumes between 1796 and 1820.[61] The relevance of the *Geschichte* to this study lies in its exemplarity for the general blend of historical and natural study in Göttingen at the time. Involved in Eichhorn's project were Johann Friedrich Gmelin (1748–1804), who occupied a chair in chemistry in Göttingen, and Johann Carl Fischer (1760–1833), who taught mathematics and physics at the University of Jena, but who had also been a history student of Gatterer.[62]

The *Geschichte* is also a relevant source because it propagated a fact-based epistemology. In an introductory volume, Eichhorn argued that "the historian should stick to [*Facta*] and let them speak for themselves ... in my opinion, any other approach is not history itself, but its use for predetermined purposes."[63] Thus, in Eichhorn's view, facts could speak for themselves, and historians should let them. Even though this attitude indicates a renewed interpretation of the role of facts as compared to Schlözer's, who encouraged historians not to let facts speak for themselves but to arrange them in unified systems, Eichhorn's words should not be taken too literally. In practice, he adhered to the same universalist spirit as his Göttingen colleagues.[64]

To sum up, I have pointed out how the term "fact" was used by the members of the Göttingen School. In Schlözer's work, a fact referred either directly to an event, such as a battle, or to a proposition based upon a range of events, such as a list of kings from a certain era. In the final decades of the eighteenth century, critically examined facts, as extracted from manuscripts, came to be considered as the basic units of proper, empirical historiography. Still, the ultimate goal of the new fact-oriented empiricism among the members of the Göttingen School remained to construct a system of facts, analogous to that of Linnaeus.

### FACTS IN LICHTENBERG'S EXPERIMENTAL *NATURLEHRE*

Parallel to the increasing popularity of the concept of fact among the members of the Göttingen School, another of the university's most distinguished scholars, Georg Christoph Lichtenberg, adopted it to express his views on the method of experimental physics.

Lichtenberg was professor of physics in Göttingen between 1770 and 1799. He was the teacher of famous German scholars such as Alexander von Humboldt (1769–1859) and Carl Friedrich Gauss (1777–1855), and the editor of four widely read and highly influential textbooks on physics published between 1784 and 1794, which were updated editions of Johann Polycarp Erxleben's (1744–1777) *Anfangsgründe der Naturlehre* (1772). According to William Clark, Lichtenberg's *Anfangsgründe* became so popular that, in the final decades of the eighteenth century, "virtually everyone who taught at a German-language university used the work," including Immanuel Kant (1724–1804).[65] As a result, Lichtenberg strongly influenced the future generation of German physicists.

Indeed, Lichtenberg was a crucial figure in establishing unity and continuity within the emerging discipline of physics in Germany.[66] Lichtenberg's textbooks and lectures updated the German learned community about the latest international advancements, but also established standards of experimentation and epistemology.[67] In 1790, Alexander von Humboldt praised him for this after attending his lectures in previous years: "I regard not simply the sum of positive knowledge that I gained from your lectures—but still more the general direction that my course of thought took under your guidance. Truth is valuable in itself, but more valuable still is the skill to find it."[68]

---

65. William Clark, "From Enlightenment to Romanticism: Lichtenberg and Göttingen Physics," in Rupke, *Göttingen Natural Science* (ref. 59), 72–85, on 74–75.

66. Claire Baldwin, "Performance and Play: Lichtenberg's Lectures on Experimental Physics," in *Performing Knowledge, 1750–1850*, ed. Mary Helen Dupree and Sean B. Franzel (Berlin: Walter de Gruyter, 2015), 193–220, on 210–11.

67. Steven Tester, "Nature, Knowledge, and Scientific Theories in GC Lichtenberg's Reflections on Physics," *HOPOS* 6, no. 2 (2016): 185–211, on 194.

On the one hand, Lichtenberg, like Gatterer and Schlözer, was a discipline builder, since he did much to increase the status of *Experimentalphysik* in German intellectual culture. On the other hand, Lichtenberg, like his colleagues, constantly moved across the not so firmly established boundaries between historical and natural study.[69] Lichtenberg had been a student of Gatterer in the 1760s, he was an active member of his Historical Institute, and he contributed to Eichhorn's *Geschichte* with an extensive biography of Copernicus.[70] Lichtenberg thus not only appreciated historical craftsmanship, but practiced it himself. His acclaimed aphorisms, collected in the famous *Sudelbücher*, contain many remarks on the proper methods of historiography. Last but not least, Lichtenberg encouraged his fellow naturalists to engage in the history of their fields, and his historical awareness developed to such an extent that he formulated a theoretical perspective on revolutions in science.[71]

Thus, Lichtenberg shared more than a university with Gatterer and Schlözer. Their scholarly enterprises crucially overlapped. Furthermore, they drew upon a similar conceptual repertoire, which came to include the concept of fact from the 1770s onward. This becomes evident from letters circulating between Lichtenberg, Schlözer, and other prominent Göttingen academics, such as Johann Friedrich Blumenbach (1752–1840). In 1797, Lichtenberg shared a recent publication with Blumenbach, which he recommended as follows: "I would not have offered it to you, had I not been assured by a reputable source that it contains only [*Thatsachen*] (or, according to Schlözer, [*Tatsachen*])."[72] Apart from showing that Göttingen academics from different fields used similar epistemological language, this fragment also indicates that there were different views on the proper spelling of *T(h)atsachen* (Schlözer's *Thatsatz* was not even mentioned here). This again illustrates the terminological complexity so characteristic of the early history of the concept of fact in the German language.

---

68. As quoted by Baldwin, "Lichtenberg's Lectures" (ref. 66), 209.

69. This has enticed some to portray Lichtenberg, somewhat anachronistically, as a "humanist-scientist": Franz H. Mautner and Franklin Miller Jr., "Remarks on GC Lichtenberg, Humanist-Scientist," *Isis* 43, no. 3 (1952): 223–31.

70. Lothar Kolmer, "G. Ch. Lichtenberg als Geschichtsschreiber," *Archiv für Kulturgeschichte* 65, no. 2 (1983): 371–416.

71. I. Bernard Cohen, *Revolution in Science* (Cambridge, MA: The Belknap Press of Harvard University Press, 1985), 225 and 517–19.

72. Ulrich Joost and Albrecht Schöne, eds., *Lichtenberg: Briefwechsel*, vol. 4 (Munich: C.H. Beck, 1992), 1025.

**Facts Versus Theories**

Like Schlözer, Lichtenberg began to use the concept of fact in his writings from the late eighteenth century onward. In his textbooks on *Naturlehre*, *Thatsachen* came to replace *Begebenheiten*.[73] Yet, facts were less prominent in Lichtenberg's work than in Schlözer's. This was true with regard to German *Naturlehre* in general. Johann Samuel Traugott Gehler's (1751–1795) popular late-eighteenth-century *Physikalisches Wörterbuch*,[74] for example, contained paragraphs on *Beobachtung*, *Erfahrung*, and *Hypothese*, but not on *Factum* or *Thatsache*. Despite the relatively minor role of the fact in this period, it is still important to examine interpretations of the concept that were prevalent then, since these predetermined those of nineteenth-century German physicists.

Lichtenberg generally considered facts as true, fixed, and impersonal inferences from multiple observations or experiences. Yet, Lichtenberg's conception of factuality was somewhat ambiguous; he never defined what exactly he meant by a fact. Like his Göttingen colleagues writing history, Lichtenberg criticized speculative approaches, while strongly advocating an empiricist epistemology.[75] Lichtenberg's use of the concept of fact fitted his empiricism. For example, he admonished his fellow scholars to be careful not to be seduced by unfounded ideas, but to be guided by facts. "Men are for the most part such reckless observers," Lichtenberg argued, "that a philosopher can never be careful enough when he wishes to use their stories of extraordinary incidents. He should never formulate any hypotheses on the basis of anything other than clear [*Thatsachen*] and experiments, if he does not want to run the risk of seeing them collapse like a house of cards."[76]

Like Schlözer, Lichtenberg thus maintained that the basis of proper *Wissenschaft* concerned the compilation of facts. The parallels between

---

73. Lichtenberg's predecessor, Erxleben, did not use the term "fact" to conceptualize the empirical basis of physics. Rather, he used the term *Naturbegebenheiten* (natural events). Johann C. P. Erxleben, *Anfangsgründe der Naturlehre* (Göttingen/Gotha, 1772), 4.

74. Johann S. T. Gehler, *Physikalisches Wörterbüch, oder, Versuch einer Erklärung der vornehmsten Begriffe und Kunstwörter der Naturlehre*, 6 vols., (Leipzig, 1787–1796).

75. Tester, "Lichtenberg's Reflections on Physics" (ref. 67), 189.

76. Wiard Hinrichs, Albert Krayer, and Horst Zehe, eds., *Lichtenbergs annotiertes Handexemplar der vierten Auflage von Johann Christian Polykarp Erxleben: "Anfangsgründe der Naturlehre"* (Göttingen: Wallstein Verlag, 2005), 801.

Lichtenberg's ideas on the methods of *Naturlehre* and Schlözer's on those of *Universal-Historie* are further evident when considering that Lichtenberg encouraged his peers not only to collect individual facts, but also to connect different facts with one another. As he explained in his lecture notes: "it is useless to find facts, when one does not try to bring them in relation to one another. We must certainly have [*Thatsachen*], since they are the basis of [*Wissenschaft*], but they are not [*Wissenschaft*] itself."[77] For Lichtenberg, thus, empirical facts were the basic elements of science, but they always should be followed by synthesis. It seems that Lichtenberg believed that this could be done without using theory.

Indeed, Lichtenberg was as devoted to observation as he was suspicious of theory. He proposed in "textbooks no longer to use the word [*Theorie*] in the sections on fire, electricity, and magnetism...but rather [*Facta*] and conjectures."[78] Lichtenberg's emphasis on the empirical basis of *Physik* can be understood in relation to his reputation of a careful and skilled experimenter.[79] When publishing about his experiments, Lichtenberg took great pains to stay true to his ideal of theory-free description. As William Clark has pointed out, Lichtenberg in his publications claimed that "his intent was not to defend any particular theory...theory-neutral, phenomenal description remained the goal of physics."[80]

An important influence on Lichtenberg's experimentalism was the Genevan naturalist and instrument-maker Jean-André Deluc (1727–1817). Lichtenberg's writings contained lots of references to Deluc's work, and the two corresponded frequently.[81] Deluc and Lichtenberg shared a passion for instruments, experiment, and for facts (Deluc, who usually wrote in French, used the word *fait*).[82] On multiple occasions, Lichtenberg declared his solidarity with Deluc's position that proper *Physik* must be grounded in empirically acquired and eternally valid facts.[83] Lichtenberg praised Deluc's work because it contained "an excellent collection of facts" and because it had nothing to do with theoretical dreams: "DeLuc's system shall be guided by undeniable [*Fackta*], while all the others are lost in dreams, the one less absurd than the other, and among which the Kantian seems to me to be the most beautiful."[84]

---

77. Albert Krayer and Thomas Nickol, eds., *Georg Christoph Lichtenberg: Vorlesungen zur Naturlehre. Notizen und Materialen zur Astronomie und Physischen Geographie* (Göttingen: Wallstein Verlag, 2013), 500.

78. As translated and quoted by Clark, "Lichtenberg and Göttingen Physics" (ref. 65), 77.

79. Mautner and Miller, "Humanist-Scientist" (ref. 69), 225.

80. Clark, "Lichtenberg and Göttingen Physics" (ref. 65), 77. Lichtenberg's "agnostic position" with regard to theory is also discussed in: Geoffrey Winthrop-Young, "Lichtenberg und die Französische Revolution: Zum Verhältnis von Sprache, Naturwissenschaft und Aufklärung" (PhD dissertation, The University of British Columbia, 1991), 105.

81. Heilbron, "Physics and Its History" (ref. 59), 61.

82. The increasing interest in and presence of measuring instruments in Germany in the late eighteenth century cannot be seen separately from parallel technological developments. For a balanced view on this issue, see: Klein, *Humboldts Preußen* (ref. 15).

Here, Lichtenberg mentions another influence on him, namely Kant. Although Lichtenberg was influenced by Kantian philosophy to a reasonable extent, their interpretation of the concept of fact was at odds. Unlike Lichtenberg, Kant thought that it was possible to extend the concept of fact from the realm of experience to the realm of reason.[85] In 1790, Kant declared: "I extend the concept of a [*Thatsache*] beyond the usual meaning of this word [because] it is not necessary, indeed not even feasible, to restrict this expression merely to actual experience."[86] Ultimately, this led Kant to define *Thatsachen* as "objects for concepts the objective reality of which can be proved . . . whether through pure reason or through experience."[87] The lack of agreement between Kant's and Lichtenberg's understanding of a fact underlines once again that different interpretations of the concept coexisted in the German context around 1800.

**Eternal Facts and Revolutions**

Albeit not Kantian, Lichtenberg's division between fact and theory was characteristic of his time. I have stressed that Lichtenberg's vocabulary resonated with the one employed by the Göttingen School. His views on factuality mirrored those of other German naturalists as well. Lichtenberg's contemporaries generally emphasized the fragile nature of a theoretical "building" (*Lehrgebäude*), while stressing the eternal value of facts. For example, Lichtenberg's former student Alexander von Humboldt wrote to Blumenbach in 1795 that "[*Thatsachen*] are fixed when the fleetingly improvised theoretical

---

83. Horst Zehe, Albert Krayer, and Wiard Hinrichs, eds., *Georg Christoph Lichtenberg: Vorlesungen zur Naturlehre. Notizen und Materialen zur Experimentalphysik, Teil 1* (Göttingen: Wallstein Verlag, 2007), 674.

84. Krayer and Nickol, *Lichtenberg: Vorlesungen zur Naturlehre* (ref. 77), 384.

85. Barbara Cassin, Emily S. Apter, Jacques Lezra, Steven Rendall, and Michael Wood, eds., *Dictionary of untranslatables: A philosophical lexicon* (Princeton, NJ: Princeton University Press, 2014), 1113–17.

86. Paul Guyer and Eric Matthews, eds., *Critique of the Power of Judgment* (Cambridge: Cambridge University Press, 2000), 332 (footnote).

87. Ibid.

[*Lehrgebäude*] has long collapsed." For this reason, Humboldt, like his teacher Lichtenberg, argued for a stringent separation between fact and conjecture: "I shall separate the [*Thatsachen*] themselves from my conjectures. This way of treating natural phenomena to me seems the most fertile and sound."[88] In Gehler's dictionary, the value of facts was stressed in terms of a building metaphor as well: "the unprejudiced naturalist must not forget that a single [*Thatsache*] has more truth value than the artificial structure of explanations."[89]

The contrast evoked between a shaky theoretical "building," which could be torn down, and eternal facts, which could not, might be interpreted as a sign of the times. The decades around 1800 were shaped by deep social and political shifts, as exemplified by the French revolution. In German *Naturlehre* around 1800, the opposition between fact and theory was mostly referred to in the context of debates about yet another revolution with French origins, namely Lavoisier's introduction of anti-phlogistic chemistry. In Göttingen, attitudes toward the new chemistry and French revolutionary politics were generally dismissive.[90]

In the 1790s, Lichtenberg often called attention to the parallels between French revolutionary chemistry and politics. As Geoffrey Winthrop-Young has argued, the French Revolution "appeared to him primarily as a war of words and signs."[91] Likewise, Lichtenberg disregarded Lavoisier's "system" for its "funny nomenclatures . . . that do not express [*Thatsachen*], but opinions."[92] Lichtenberg's friend Deluc, who resented French science in general, even went as far as to claim that the objectionable French political revolution had its roots in the new theoretical system of anti-phlogistic chemistry. On November 18, 1793, he wrote to Lichtenberg that "this French quest for domination began, I say, with chemistry. There we must demolish their edifice, because there we can demonstrate . . . the truth."[93]

With the aversion toward the French came an increased appreciation of English natural philosophy, particularly the kind represented by Francis Bacon (1561–1626). In Bacon, Lichtenberg and his fellow empiricists found a common enemy of speculative systems and theories, and a prominent proponent of empirical facts.[94] In Johann Fischer's *Geschichte der Physik*, for example, Bacon was portrayed as "the first to show the true way," whereas Descartes and his followers were blamed for "wander[ing] from the true path of experience."[95] Alexander von Humboldt propagated Bacon as the one and only founder of modern experimentalism, as well. According to historian Michael Dettelbach, Humboldt's turn to Baconianism in the late eighteenth century was motivated by his "hostility toward theory," and his "faith in the positivity and independence of mere facts."[96] Given Bacon's crucial role in establishing the concept of fact in English natural philosophy,[97] the establishment of this English-German connection in the final decades of the eighteenth century cannot be seen separately from the increasing momentum of the concept of fact in the German context.[98]

---

88. Alexander von Humboldt, "Ueber die gereizte Muskelfaser, aus einem Briefe an Herrn Hofrath Blumenbach," *Neues Journal der Physik* 2 (1795): 115–29, on 127.
89. Gehler, *Physikalisches Wörterbuch* (ref. 74), vol. 2, 678.
90. Heilbron, "Physics and Its History" (ref. 59), 61–65.
91. Winthrop-Young, "Lichtenberg und die Revolution" (ref. 80), ii.
92. As quoted in ibid., 137.
93. Heilbron, "Physics and Its History" (ref. 59), 65.

Apart from the English influence, it is likely that the crumbling status of phlogiston theory, as well as the historical projects Lichtenberg and his Göttingen naturalist colleagues were involved in, made them very much aware of the temporary character of theories. Acknowledging these backgrounds, we can appreciate their desire for lasting, empirical knowledge. Hence, German naturalists like Lichtenberg, Gehler, and Humboldt stressed the eternal character of facts, and opposed them to temporary and fallible theoretical "buildings" and hypotheses. They regarded factual knowledge as true and eternal and in contrast to theory.[99]

Did similar reasons motivate the embrace of facts by Schlözer and the other members of the Göttingen School? It is certain that they shared Lichtenberg's aversion to nonfactual systems. Like Lichtenberg, they considered such systems characteristic of French knowledge culture in particular. Indeed, the members of the Göttingen School shared Lichtenberg's rejection of French scholarship and his distaste for revolutions, particularly the French.[100] It thus seems plausible that Schlözer's emphasis on facts, like Lichtenberg's, related to the political context of the French Revolution. That said, the advent of antiphlogistic chemistry seems not to have played a direct role in the Göttingen School's embrace of factuality, despite the commitment of the members of this school to natural study. First and foremost, the parallel emergence of a culture of fact in *Universal-Historie* and *Naturlehre* in Göttingen was part of a common struggle against hitherto dominant, rationalist methods.

---

94. Clark, "Lichtenberg and Göttingen Physics" (ref. 65), 80.

95. As quoted by Heilbron, "Physics and Its History" (ref. 59), 59.

96. Michael Dettelbach, "'Baconianism' in Revolutionary Germany: Humboldt's 'Great Instauration'," in *The Skeptical Tradition Around 1800*, ed. Johan van der Zande and Richard H. Popkin (Dordrecht: Springer, 1998), 175–86, on 178.

97. Shapiro, *Culture of Fact* (ref. 3); Daston, "Baconsche Tatsachen" (ref. 23).

98. David Simpson has explored the links between what he calls the English "revolt against theory" and contemporary nationalist discourse in the wake of the French revolution. David Simpson, *Romanticism, Nationalism, and the Revolt Against Theory* (Chicago: The University of Chicago Press, 1993).

99. A study of German chemistry textbooks around 1800 underlines this interpretation of the of fact as eternal: Jan Frercks and Michael Markert, "The Invention of Theoretische Chemie: Forms and Uses of German Chemistry Textbooks, 1775–1820," *Ambix* 54, no. 2 (2007): 146–71, on 155. Also Frercks, "Facticity of Facts" (ref. 3).

### BEFORE DISCIPLINES: THE FACT AS A HISTORICAL CONCEPT

The functions and increasing importance of the concept of fact in late-eighteenth-century Göttingen are better understood when seen in relation to the then-dominant classifications of knowledge. I have shown that although dynamics of differentiation developed in this context, it would be mistaken to suppose that there were strict disciplinary boundaries. Indeed, historical and natural study remained intertwined in the ideals and practices of Göttingen scholars. In fact, the boundaries between historical and natural research were still blurred in Germany generally.

This conformed to the classification of knowledge propagated by Christian Wolff, which remained dominant in German intellectual culture far into the eighteenth century. Wolff and his many followers did not consider "history" as an independent field of knowledge, but as providing the empirical basis for rational science. In their model, history comprised both human and natural knowledge.[101] This scheme was still endorsed by Kant, who characterized experimental physics as "historical," and attributed to it only a precursory role in rational, mathematical knowledge. For Kant, the value of historical or empirical knowledge lay in its creation of "the objects of rational knowledge."[102] According to Wolff and Kant, historical or empirical knowledge was thus positioned hierarchically below mathematics and philosophy.[103] For a significant part of the eighteenth century, the historical and the empirical were thus understood equivalent. But they were not regarded as *wissenschaftlich*. Only philosophy and mathematics were considered *Wissenschaft*, as the methods of both enabled explanation. History, on the other hand, merely had to do with the recognition, description, and ordering of individual events and observations.

---

100. Reill, *Rise of Historicism* (ref. 39), 8.

101. Arno Seifert, *Cognitio Historica: Die Geschichte als Namengeberin der frühneuzeitlichen Empirie* (Berlin: Duncker & Humboldt, 1976), 163–78. Gianna Pomata and Nancy G. Siraisi, eds., *Historia: Empiricism and Erudition in Early Modern Europe* (Cambridge, MA: The MIT Press, 2005).

102. As quoted by Seifert, *Cognitio Historica* (ref. 101), 185–86.

I have argued above that the objects of historical knowledge were increasingly often labelled as facts (*Facta*, *Thatsachen*, or *Tatsätze*) rather than events (*Begebenheiten*) toward the end of the eighteenth century. In German learned culture, facts appeared regularly first in texts on the foundations of theological history, somewhat later in the Göttingen tradition of *Universal-Historie*, and not long thereafter in the historical subject of *Experimentalphysik*. In late-eighteenth-century German intellectual culture, the concept of fact thus developed as part of a historical repertoire, which comprised both human and natural study.

As historical knowledge was considered hierarchically below philosophy and mathematics, facts were initially considered only the basis of *Wissenschaft*, but not *Wissenschaft* itself. As I have shown above, both Schlözer and Lichtenberg argued along these lines. These commonalities illustrate the common conceptual foundations of German physics and historiography, which would develop into modern academic disciplines over the course of the nineteenth century.

Evidently, the hierarchical classificatory scheme outlined above differed fundamentally from the disciplinary classification that emerged in the nineteenth century. As Rudolf Stichweh has pointed out, the modern disciplinary system contained relatively autonomous units rather than hierarchical layers.[104] The evolving disciplinary landscape was "horizontal": mathematics and philosophy came to stand on the same hierarchical level as fields of study that had previously been considered "historical," including physics and historiography. The context in which this transformation took place was the expansion of the philosophical faculties at German universities, many of which adopted the Göttingen model of specialization in the decades around 1800. The structural organization of specialized training in these expanding institutional contexts was a key aspect of the materialization of physics and history as disciplines.[105] Through practical exercises, for example, disciplinary methods

---

were established, refined, and passed on to next generations. Furthermore, the boundaries of disciplines became more clearly established by the foundation of institutes, journals, and university chairs.

In this mutating institutional landscape, the concept of fact mutated as well. As the status of grand theories and systems further crumbled, the status of facts further increased. After 1800, facts were no longer seen as providing merely the basis of *Wissenschaft*. Both physicists and historians began to grant the facts themselves the status of *Wissenschaft*.

## FACTS AND DISCIPLINE FORMATION

In this section, I trace the history of the concept of fact in contexts of discipline formation after 1800. I show how facts acquired prominence in physics and historiography, and how they related to specific teaching and research practices that came to characterize these disciplines. My main focus will be on the University of Berlin, which opened in 1810 and was modelled after the institutional structures of the University of Göttingen.[106] Whereas a German culture of fact first gained a foothold in late-eighteenth-century Göttingen, it became genuinely visible in early-nineteenth-century Berlin. Here, an intellectual battle raged between defenders of philosophical, speculative methods on the one hand, and proponents of empirical, fact-oriented ones on the other. Simultaneous to these debates, processes of differentiation and specialization intensified, which sharpened the contours of physics and history as disciplines.

### The Growing Contrast Between Fact and Theory in German Physics

In the first decades after 1800, physics and chemistry remained closely connected.[107] Sometimes, both were still regarded as part of *Naturlehre*. The general trend, however, was that chemistry differentiated from *Naturlehre* as a separate field of study, which made physics something of a residual discipline. Consequently, physics, which was now often perceived of as a synonym of

---

106. A lot has been written on the foundation of the University of Berlin, its indebtedness to Göttingen, and the role of these universities in the formation of modern scientific and humanistic disciplines. See, e.g.: Stichweh, *Sociology of Scientific Disciplines* (ref. 35), 9; Chad Wellmon, *Organizing Enlightenment: Information Overload and the Invention of the Modern Research University* (Baltimore: Johns Hopkins University Press, 2015), 210–33.

107. Stichweh, *Zur Entstehung wissenschaftlicher Disziplinen* (ref. 11), 94–172.

*Naturlehre*, took more of a distinctive shape in the first two decades of the nineteenth century. In this period, physicists defined their materializing discipline as a pursuit of facts. A clear-cut example of a physicist employing this demarcation strategy was Ernst Gottfried Fischer (1754–1831).

Fischer, who had been a private tutor of the Humboldt brothers, became professor of physics at the University of Berlin right at its foundation in 1810.[108] In his *Lehrbuch der mechanischen Naturlehre*, which first appeared in 1805, Fischer aimed to establish physics, which he labelled *mechanische Naturlehre*, as a mathematical and empirical discipline based on mechanics.[109] Fischer unequivocally embraced the notion of fact in his characterization of physics and its proper method. In the preface to the second edition of his book, Fischer accused authors of previously published German textbooks on physics, including Gehler, of having mixed up facts and hypotheses in their discussion of Lavoisier's theory. Subsequently, Fischer defined facts not as the eternal foundation of an ever-changing scientific *Lehrgebäude*, which would have been a common expression in the late eighteenth century, but as the *Lehrgebäude* itself: "A well-ordered [*Lehrgebäude*] of an empirical science," Fischer stressed, "can be nothing else than a careful, systematic compilation of [*Thatsachen*]." As long as knowledge was grounded in fact, Fischer argued, it could stand the test of time: "Such a building is established in its essential parts for eternity. Discoveries can serve to enlarge it or to improve its inner configuration, but never to destroy it and tear it down."[110] For Fischer, facts were thus empirical, eternal, and unchangeable building blocks that, in differing combinations, constituted scientific (*wissenschaftliches*) knowledge. More precisely, they were generally valid propositions directly based on empirical observations. For example, Fischer regarded it "a very common [*Thatsache*] that every chemical mixture of two substances produces either heat or cold."[111] Furthermore, he found it "a well-known [*Thatsache*]" that "some bodies absorb light, and in the dark emit it again."[112]

---

108. Klein, *Humboldts Preußen* (ref. 15), 279. On the political backgrounds of the foundation of the University of Berlin: e.g., Charles E. McClelland, *State, Society, and University in Germany, 1700–1914* (Cambridge: Cambridge University Press, 1980), 101–50.

109. References are to the third edition of Fischer's book from 1827, which includes the preface to the second edition from 1819. These editions of Fischer's two-volume book were aimed to be used in physics instruction at German universities and became very influential. Ernst G. Fischer, *Lehrbuch der mechanischen Naturlehre*, 2 vols. (Berlin/Leipzig: G.C. Nauck, 1827).

110. Ibid., vol. 1, xi.
111. Ibid., vol. 1, 208.
112. Ibid., vol. 2., 131.

Fischer's emphasis on facts must be understood as a response to certain trends that he perceived as threatening his emerging discipline.[113] To begin with, Fischer criticized the "dark, unclear, and mystical concepts" of Johann Wolfgang von Goethe's (1775–1854) anti-Newtonian theory of color, which Goethe had published in 1810.[114] He disregarded Goethe's theory, because it failed to relate to mathematics and experiment. According to Fischer, Goethe ignored the facts. Fischer renounced the contemporary German movement of *Naturphilosophie* for similar reasons. He lamented that it "must hide behind obscure and mystical imaginations everywhere, in order to avert the attacks from the realm of experience which can be made against it."[115]

One of Fischer's younger colleagues in Berlin, the experimentalist Heinrich Gustav Magnus (1802–1870), despised the speculative methods of *Naturphilosophie* as well. On November 2, 1831, twelve days before the death of the prominent idealistic philosopher Georg Wilhelm Friedrich Hegel (1770–1831), Magnus wrote that he hoped that "*Naturphilosophie* will go to the grave together with Hegel."[116] Magnus, like Fischer, proclaimed that proper physics should be strictly empirical and non-speculative.[117] He spread this faith, for example, by organizing small-scale practical training in his private physical laboratory in Berlin.[118] In 1870, one of Magnus' many famous pupils, Hermann von Helmholtz (1821–1894), characterized him as an ardent opponent of speculation, and a devotee of facts: "Wherever he came, he brought forth an abundance of new and often surprising [*Thatsachen*], which he had carefully and responsibly observed, and integrated in the great building of *Wissenschaft*."[119]

Thus, both Magnus and Fischer employed the concept of fact to define the boundaries of their emerging discipline. They believed that physics comprised nothing but a constellation of well-established and eternal facts. For Fischer and Magnus, facts rather than theories had become the core of *Wissenschaft*. Such definitions of empirical facts as having value for their own sake appeared especially from the early nineteenth century onward. These findings endorse Daston's assertion that, around 1800, the category of fact underwent a profound transformation. She has argued that "eighteenth-century savants had revered facts but had believed them to be the alpha, not the omega, of scientific achievement," whereas nineteenth century scientists held pure facts to be "the last, best hope for permanence in scientitic achievement."[120] Based on the above, one might add to Daston's analysis that, even though such opinions were indeed rare among German savants before 1800, their interpretation of facts as eternal and in opposition to theory was a precondition for the enhanced status of facts in the early nineteenth century.

---

113. Dennis L. Sepper, *Goethe Contra Newton* (Cambridge: Cambridge University Press, 1988), 158–73.
114. Fischer, *Lehrbuch* (ref. 109), vol. 1, xii.
115. Ibid., vol. 1, xiv.
116. Edvard Hjelt, *Aus Jac. Berzelius' und Gustav Magnus' Briefwechsel in den Jahren 1828–1847* (Braunschweig: Friedrich Vieweg und Sohn, 1900), 63.
117. H. Gustav Magnus, *Festrede auf der Universität zu Berlin am 3. August 1862 gehalten* (Berlin: Königliche Akademie der Wissenschaften, 1862), 24.
118. On the details of Magnus' laboratory training and its significance for the development of German physics: Dieter Hoffmann, ed., *Gustav Magnus und sein Haus* (Stuttgart: GNT-Verlag, 1995).
119. Hermann von Helmholtz, "zum Gedächtnis an Gustav Magnus," in *Vorträge und Reden* (Braunschweig: F. Vieweg und Sohn, 1884), vol. 2, 33–51, on 49.

While Berlin physicists Fischer and Magnus demarcated their field from philosophy, the ties between physics and mathematics were becoming stronger. The mathematization of German physics over the course of the nineteenth century further anchored the opposition between fact and theory. The Königsberg seminar for physics, which was founded in 1834 by Franz Ernst Neumann (1798–1895), played a leading role in this development.[121]

Inspired by astronomer Friedrich Wilhelm Bessel's (1784–1846) work on probability calculus, Neumann trained his seminar students to balance between mathematical theory and experiment by teaching them how to quantify measurement error.[122] Bessel's epistemology influenced Neumann as well. In Bessel's view, experiment-based mathematical theories in physics could never be true, but only probable. In an 1838 lecture on probability calculus, Bessel claimed that "any science inferring from experience to theory begins with observations, then learns from probability theory in order to employ and to utilize these observations, and in that way eventually derives the most probable theory."[123] Bessel thus insisted that physical theories and experimental results should never be considered true. But he stuck to a definition of facts as true and unchangeable, equating them with mathematical truths.[124] Neumann and his seminar students embraced Bessel's insight that experimental error was fundamentally unavoidable. Hence, they maintained that, unlike in pure mathematics, absolute truth could never be fully achieved in experimental settings.[125] In 1850, Neumann declared he aimed to teach students in his seminar how "to identify in the physical [*Thatsachen*] those aspects that allow for mathematical treatment."[126] The truth-value of these facts remained beyond question, but the results of their mathematical treatment, which was considered to take place on the level of theory, contained only probabilistic truth. Bessel and Neumann's distinction between probable theory and true fact can be interpreted as a continuation of the tendency to define fact and theory in opposite terms, which, as I have argued, had its roots in the late eighteenth century.

---

120. Lorraine Daston, "Fear and Loathing of the Imagination in Science," *Daedelus* 127, no. 1 (1998): 73–95, on 90.

121. Kathryn M. Olesko, *Physics as a Calling: Discipline and Practice in the Königsberg Seminar for Physics* (Ithaca, NY: Cornell University Press, 1991); Stichweh, *Zur Entstehung wissenschaftlicher Disziplinen* (ref. 11), 367–70.

122. Neumann recommended Ernst Fischer's textbook as preparatory reading material for his seminar. Olesko, *Physics as a Calling* (ref. 121), 156–57.

123. Friedrich W. Bessel, "Ueber Wahrscheinlichkeits-Rechnung," in *Populäre Vorlesungen über wissenschaftliche Gegenstände*, ed. Heinrich C. Schumacher (Hamburg: Perthes-Besser & Mauke, 1848), 387–407, on 406.

### The Incomplete and Objective Facts of *Quellenkritik*

I have shown that the concept of fact played a prominent role in the shaping of the identity of a German discipline of physics, and that German physicists linked the pursuit of fact to practices such as laboratory work and mathematical precision measurement, which they taught at specialized university institutes. In the remaining part of this section, I shift focus to the emerging historical discipline, discussing two of its foundational figures: Wilhelm von Humboldt (1767–1835) and Leopold von Ranke (1795–1886). Humboldt and Ranke interpreted and used the concept of fact in different ways. Furthermore, their understanding of facts differed with respect to their Göttingen predecessors and their contemporaries in physics.

Although no practitioner of history himself, Wilhelm von Humboldt was deeply concerned about the foundations of the emerging discipline. As a disciple of Schlözer and an admirer of the Göttingen School in general, he advocated the relevance of philological methods for historiography.[127] Still, his views on the foundations of historiography deviated from those of his teacher Schlözer in multiple respects. The fluid concept of fact, to which Humboldt added his own interpretation, lays bare some of these differences. While putting forward a historical methodology that balanced reason and experience, Humboldt adjusted the notion of fact. He defined it as the creative and imperfect result of historical research rather than as its unquestionable empirical basis.

---

124. Ibid., 387.
125. Kathryn M. Olesko, "The Meaning of Precision: The Exact Sensibility in Early Nineteenth-Century Germany," in *The Values of Precision*, ed. M. Norton Wise (Princeton, NJ: Princeton University Press, 1995), 103–34, on 115; Olesko, *Physics as a Calling* (ref. 121), 451–58.
126. As quoted by Stichweh, *Zur Entstehung wissenschaftlicher Disziplinen* (ref. 11), 223.
127. Beiser, *Historicist Tradition* (ref. 13), 170–71.

Humboldt shared his views on proper historical method in a famous 1821 lecture called "*Ueber die Aufgabe des Geschichtschreibers.*"[128] At the Prussian Academy, Humboldt claimed it to be the basic task of the historian "to show what has happened." But he readily acknowledged this task to be impossible, because of the fundamental incompleteness of historical source material. This had implications for the results of history writing, which Humboldt labeled "the facts of history." As he explained it: "the [*Thatsachen*] of history are little more than the results of tradition and research, which one has come to accept as true."[129] Furthermore, Humboldt argued that with the best-as-possible reconstruction of what has happened—that is, by the establishment of historical facts—the historian had fulfilled only one part of his task. Humboldt emphasized that the historian should aim to capture the true historical facts in their entirety. "The truth of anything that has happened relies on the inclusion of the invisible part of every [*Thatsache*]," he argued.[130] Humboldt thus urged historians to complete historical facts. Proper historiography meant unearthing the ideas behind the mere manifestation of events. This practice of unearthing was necessary because of the fundamentally incomplete rendition of the past in historical sources. Only in this way, Humboldt claimed, could historical truth be approximated.

For Humboldt's Enlightenment predecessors, such as his mentor Schlözer, the aim of historiography had been to collect self-evident facts and to arrange them in systems. Although Humboldt appropriated the Göttingen School's emphasis on facts, his aims were different. Rather than to arrange them in a system, he incited historians to represent the historical facts themselves as completely as possible. This he understood to be a creative process in which the imagination of the historian played an important role.

---

128. Wilhelm von Humboldt, "Ueber die Aufgabe des Geschichtschreibers," in *Gesammelte Schriften*, ed. Albert Leitzmann, 13 vols. (Berlin: Behr, 1905), vol. 4, 35–56.
129. Ibid., 36.
130. Ibid.

The fruition of the German historical discipline has probably been associated most emphatically with Ranke. His adage that the historian ought to show "how it actually was" is notorious.[131] Furthermore, it is well known that Ranke attached importance to facts. Ranke encouraged his fellow historians not to ignore the facts, even if they were "*unschön*."[132] In an 1831 letter to his brother Heinrich, Ranke stated that "my basic idea is to recognize, to penetrate, and to show the [*Fakten*] as they are. The true doctrine lies in the knowledge of [*Thatsachen*]."[133]

By now it may be clear that Ranke was not the first to put facts at the heart of German historical research. Indeed, Ranke's found much inspiration in the empirically oriented and anti-speculative ideals of the Göttingen School.[134] The impact of Ranke's appeal to factual method, however, was even greater than that of his Göttingen predecessors had been. This was due especially to Ranke's teaching at the University of Berlin between 1825 and 1870. Parallel to the pedagogical innovations in physics of Magnus and Neumann in Berlin and Königsberg, Ranke's small-scale and method-oriented "exercises" (*Übungen*) were of great importance to the formation of the historical discipline. Like his colleagues teaching physics, Ranke emphasized that his teaching focused on facts. At one point, Ranke recalled it as the goal of his earliest exercise sessions in Berlin to have been "the establishment of [*Thatsachen*] in chronological order."[135]

Moreover, Ranke shared Magnus' aversion to philosophical methods. In the 1820s, he engaged in a polemic with Berlin's philosophical school of history revolving around Hegel. Hegel blamed Ranke's *Quellenkritik* for merely presenting detail after detail, without any unified picture in sight. Ranke, in turn, criticized Hegel's philosophical approach for ignoring the limitations of historical knowledge, and for deductively imposing theoretical schemes on the facts of history.[136] In an 1831 lecture, he argued that "speculation departs from the reality of the [*Thatsache*] in every aspect . . . there is an inner divide between the concept of philosophy and the [*Faktum*]."[137]

---

131. Anthony Grafton, *The Footnote: A Curious History* (Cambridge, MA: Harvard University Press, 1999), 34–61.

132. Leopold von Ranke, *Geschichten der romanischen und germanischen Völker von 1494 bis 1535* (Leipzig: G. Reimer, 1824), vii.

133. Walther Peter Fuchs, ed., *Leopold von Ranke: Das Briefwerk* (Hamburg: Hoffmann und Campe, 1949), 249.

134. Kasper R. Eskildsen, "Leopold Ranke's Archival Turn: Location and Evidence in Modern Historiography," *Modern Intellectual History* 5, no. 3 (2008): 425–53, on 431–32.

135. Kasper R. Eskildsen, "Private Übungen und verkörpertes Wissen: Zur Unterrichtspraxis der Geschichtswissenschaft im 19. Jahrhundert," in *Akademische Wissenskulturen: Praktiken des Lehrens und Forschens vom Mittelalter bis zur Moderne*, ed. Martin Kintzinger and Sita Steckel (Basel: Schwabe Verlag, 2015), 143–61, on 149.

Ranke subscribed to the Humboldtian view that the historian should make the invisible aspects of historical facts visible. During a lecture at the University of Berlin in 1845/1846, Ranke introduced the "strict method" of history as "the extraction of pure [*Faktum*]."[138] At the same time, he encouraged his students to look beyond pure fact. This was because he was "convinced that this [*Faktum*] has a spiritual content. The [*Thatsache*] is not the outermost limit. What has happened is not the ultimate thing we can know. There is something that happens inside of it."[139] So, like Humboldt, Ranke believed that there was more to historical practice than the mere establishment of facts. That being said, Ranke, like the members of the Göttingen School, seemed optimistic about the possibility of finding "pure fact" in the first place.

A novelty in Ranke's conception of factuality was the link he established between fact and "objectivity." For Ranke, objective knowledge meant uninterpreted knowledge, as directly extracted from the archive.[140] In 1862, he put it as follows: "It is important to free contemporary history from the influence of personal interests and opinions and to obtain an independent position from which a general view, the objective truth, unfolds."[141] According to Ranke, historical facts were to be presented and discussed objectively, that is, without being obscured by tradition or individual interpretation.

When comparing Ranke's ideals to his practice, his procedures strike as more speculative and less impersonal than some of his methodological statements suggest. Yet, an image of Rankean historiography as demanding nothing but objective facts, which Ranke had cultivated himself, became increasingly persistent as the nineteenth century ran its course. For example, Hermann Wesendonck, reflecting on the nature of the German historical discipline in 1876, characterized the dominant school of Ranke as wishing "to let the [*Thatsachen*] speak alone; the reader shouldn't notice one single aspect of the historian."[142] Above all, Rankean historiography was thus labelled as demanding the facts to speak for themselves, despite Ranke's repeated insistence that historians should capture the idea beyond the facts.[143]

---

136. Beiser, *Historicist Tradition* (ref. 13), 258–66.
137. Volker Dotterweich and Walther Peter Fuchs, eds., *Leopold von Ranke: Aus Werk und Nachlass, Vorlesungseinleitungen* (Munich: Oldenbourg, 1975), 75.
138. Ibid., 177.
139. Ibid.
140. Eskildsen, "Ranke's Archival Turn" (ref. 134), 437.
141. Dotterweich and Fuchs, *Ranke: Vorlesungseinleitungen* (ref. 137), 306.

**Divergent Interpretations on a Common Basis**

During the first decades of the nineteenth century, physics and history formed around different objects of research, developed distinct research practices, and founded independent institutions. Still, in this period of discipline formation, there was an epistemological discourse that transcended the emerging boundaries of physics and historiography. Both physicists and historians aimed to ban philosophical speculations from their discipline, while developing research and teaching practices that focused on facts. That being said, the practices as well as their exact interpretations of the concept of fact were different. For instance, the idea that facts needed to be completed was characteristic of German historiography, whereas the connection of facts with a probabilistic notion of truth was typical of German physics. So, although nineteenth-century German historiography and physics shared an empirical outlook and a strong orientation of facts, divergent interpretations of the concept of fact emerged on that shared basis.[144]

Until around 1800, the empirical study of the human and natural world was regarded as part of the same historical and fact-oriented knowledge tradition. For example, even though universal historian Schlözer acknowledged the differences between natural and human objects of study, he emphasized the principal unity of the methods of human and natural history. In contrast, over the course of the nineteenth century, the methods of natural and historical knowledge were portrayed increasingly often as fundamentally distinct, despite their common orientation on facts. It demands further study if and how the emergence of different conceptions of factuality in different disciplines shaped the historical demarcation between natural and historical knowledge and, ultimately, between the sciences and the humanities.[145] It is quite possible that the spiritual content and incomplete character of historical facts emphasized by Ranke and Humboldt played a role in this development, given that such interpretations were uncommon among naturalists.

---

142. Hermann F. Wesendonck, *Die Begründung der neueren Deutschen Geschichtsschreibung durch Gatterer und Schlözer* (Leipzig: J.W. Krüger, 1876), 128.

143. Later in the nineteenth century, the feasibility of uninterpreted historical knowledge and the notion of "objective fact" was severely criticized, for example, by Johann Droysen. Droysen regarded it as essential to abandon some of the "confusing concepts of *Quellenkritik*," including the "objective fact." Rudolf Hübner, ed., *Historik: Vorlesungen über Enzyklopädie und Methodologie der Geschichte* (Munich: Oldenbourg, 1937), 139.

144. Taking physics and historiography as representative disciplines for the sciences and humanities, the results of this study shed light on Julian Hamann's claim that German intellectuals defined the *Geisteswissenschaften* and the *Naturwissenschaften* as "equal but different." Julian Hamann, "Boundary Work between Two Cultures: Demarcating the Modern Geisteswissenschaften," *History of Humanities* 3, no. 1 (2018): 27–38.

## CONCLUSION

In this article, I have sketched a long-term historical trajectory of the concept of fact in the German academic context. I started from the introduction of the term *Thatsache* in 1756, and followed the concept up to the crucial role it played in the formation of the modern disciplines of physics and history. While the relations between existing fields of knowledge transformed by the genesis of the modern system of knowledge disciplines, the concept of fact transformed as well. In Germany between 1750 and 1850, it knew different interpretations in different intellectual contexts.

These varying shapes of the concept of fact—I have counted at least five pairs of contrasting interpretations—prove that the concept was genuinely fluid in the German context around 1800. To begin with, facts were usually interpreted as true, as in the Göttingen School's universal history, but also as probable, as in theological history. Furthermore, some German scholars, like Lichtenberg, considered facts to be strictly empirical. Others, however, such as Kant, argued that *Thatsachen* could be empirical *and* rational. Additionally, facts could be defined either as the object or the result of research, as speaking for themselves or requiring completion, and as directly referring to events or phenomena or as propositions about events or phenomena. These ambiguities surrounding the concept of fact in the studied contexts were further magnified by the circulation of multiple terms and spellings, including *factum*, *Factum*, *Faktum*, *T(h)atsache*, and *T(h)atsatz*, as well as their plurals.

Despite this conceptual and terminological fluidity, it is possible to discern some general transformations of the concept of fact in the German context around 1800. These related to the formation of the modern disciplinary system at German universities. Focusing on the academic context of Göttingen, I have argued that, in the late eighteenth century, a fact-based epistemology emerged in several interrelated empirical or so-called historical fields of knowledge, comprising both human and natural study. Facts were regarded as the eternal, unchangeable, and empirical building blocks of the *Lehrgebäude* of *Wissenschaft*, and put in contradistinction with fleeting theories, hypotheses, and speculations. Initially, the fact thus existed as a historical concept. There was consensus among German scholars that, although they formed its basis, facts were not *Wissenschaft*.

---

145. Denise Phillips has studied these dynamics by tracing the changing meaning of "positive knowledge" in the first half of the nineteenth century. Denise Phillips, "Trading Epistemological Insults: 'Positive Knowledge' and Natural Science in Germany, 1800–1850," in *The Worlds of Positivism: A Global Intellectual History, 1770–1930*, ed. Johannes Feichtinger, Franz. L. Fillafer, and Jan Surman (Cham: Palgrave MacMillan, 2018), 137–53.

From around 1800 onward, this changed. For many early-nineteenth-century historians and physicists, facts, as extracted from archival or experimental study, came to constitute the essence of *wissenschaftliches* knowledge. Simultaneously, physics and history gained prestige as knowledge enterprises worth pursuing for their own sake, rather than ones put to the service of higher forms of knowledge. The new status of facts coincided with the development of discipline-specific, fact-oriented research and teaching practices. I have shown how factuality in German historiography stood central to the method of archival source criticism, as practiced in Ranke's historical exercises in Berlin, and how the familiarization of a new generation of mathematically oriented German physicists with error analysis and probability theory widened the schism between fact and theory.

Looking beyond the confines of this study, it should yet be explored if and how the concept of fact further transformed in the disciplinary vocabularies of physicists and historians in the nineteenth century. Furthermore, there remains much to discover about the history of the concept of fact in the social and behavioral sciences. Last but not least, it is worth asking how this German narrative compares to other national contexts, such as the English and the French.

I raise these questions to emphasize that this study is part of a larger history of the concept of fact, and of epistemological concepts in general. All too often, fundamental concepts like "fact," "data," or "objectivity" are taken for granted. Yet their histories, which reach across geographical and disciplinary boundaries, teach us that they have always had a flexible nature. Therefore, not these concepts themselves, but their ongoing transformations are inevitable.

### ACKNOWLEDGEMENTS

I would like to thank Andre Wakefield and one anonymous referee for their referee reports, and the editors of *Historical Studies in the Natural Sciences*, especially Jutta Schickore and Michael Gordin, for their guidance during the refereeing process. For critical comments on earlier versions of this article, I thank Jeroen van Dongen, Bart Karstens, Tom Kayzel, Fabian Krämer, Emma Mojet, Josephine Musil-Gutsch, Kathryn Olesko, Jaco de Swart, and Chaokang Tai. In addition, I am indebted to Rens Bod, Lorraine Daston, Omar Nasim, Daniel Rosenberg, and Paul Ziche for insightful discussions. This study is part of the project "The Flow of Cognitive Goods: Towards a Post-Disciplinary Perspective on the History of Knowledge," which runs between 2016 and 2020 at the University of Amsterdam, and which is funded by NWO (Netherlands Organisation for Scientific Research).

especially Jutta Schickore and Michael Gordin, for their guidance during the refereeing process. For critical comments on earlier versions of this article, I thank Jeroen van Dongen, Bart Karstens, Tom Kayzel, Fabian Krämer, Emma Mojet, Josephine Musil-Gutsch, Kathryn Olesko, Jaco de Swart, and Chaokang Tai. In addition, I am indebted to Rens Bod, Lorraine Daston, Omar Nasim, Daniel Rosenberg, and Paul Ziche for insightful discussions. This study is part of the project "The Flow of Cognitive Goods: Towards a Post-Disciplinary Perspective on the History of Knowledge," which runs between 2016 and 2020 at the University of Amsterdam, and which is funded by NWO (Netherlands Organisation for Scientific Research).